\documentclass[preprint]{aastex62}
\usepackage{epsf}
\usepackage{natbib}
\usepackage{graphicx}
\usepackage{float}
\usepackage{afterpage}
\usepackage{amssymb}
\usepackage{enumitem}
\usepackage{bm,bbm,amssymb, amsmath}
\usepackage{url}
\usepackage[section]{placeins}

\usepackage{mathrsfs}
\usepackage{ulem}

\usepackage{hyperref}

\begin{document}
\title{\Large Planetesimal drift in eccentric disks: possible outward migration}
\correspondingauthor{Kedron Silsbee}
\email{kpsilsbee@utep.edu}

\author[0000-0003-1572-0505]{Kedron Silsbee}
\affil{University of Texas at El Paso, El Paso, TX, 79968}
\affil{Max-Planck-Institut f\"ur
Extraterrestrische Physik, 85748 Garching, Germany}
\begin{abstract}
Radial drift of solid particles in the protoplanetary disk is often invoked as a threat to planet formation, as it removes solid material from the disk before it can be assembled into planets. 
 However, it may also concentrate solids at particular locations in the disk, thus accelerating the coagulation process.  Planetesimals are thought to drift much faster in an eccentric disk, due to their higher velocities with respect to the gas, but their drift rate has only been calculated using approximate means.  In this work, we show that in some cases, previous estimates of the drift rate, based on a modification of the results for an axisymmetric disk, are highly inaccurate.  In particular, we find that under some easily realized circumstances, planetesimals may drift outwards, rather than inwards.  This results in the existence of radii in the disk that act as stable attractors of planetesimals.  We show that this can lead to a local enhancement of more than an order of magnitude in the surface density of planetesimals, even when a wide dispersion of planetesimal size is considered.
\end{abstract}
\section{Introduction}

\par
Rapid radial drift of solid particles in protoplanetary disks has been identified as a major barrier to planet formation because it leads to high collision velocities and depletion of solid material in the disk.  Often referred to as the ``meter-size barrier", radial drift affects larger objects as well, with typical disk models yielding drift timescales under one Myr for objects of hundreds of meters in size at one AU \citep{Adachi76, Weidenschilling77}.  
\par
In the presence of a nearby binary companion, objects of tens or even hundreds of kilometers in size experience substantial radial drift because the binary companion excites eccentricity in both the disk and the planetesimal.  \citet{Rafikov15a} estimated that the presence of a companion star on an eccentric orbit at 20 AU increases the inspiral rate at 1 AU (for bodies larger than a few kilometers) by factors of 10's to 100's compared with the case of a circular disk. However, this estimate was made in a rather approximate way, by using the formulae derived in \citet{Adachi76} for an axisymmetric disk and replacing the planetesimal eccentricity with the relative eccentricity between planetesimal and gas.  
\par
The planet formation rate is known to be suppressed around stars that have a companion within $\sim 200$ AU \citep{Kraus16, Moe21}.  Multiple factors may lead to this suppression.  Gravitational perturbations from the companion stir up the planetesimal population, potentially resulting in destructive collisions \citep[e.g.][]{Heppenheimer78, Paardekooper08, Rafikov15b}.  In addition, the masses and lifetimes of circumstellar disks are generally reduced in tight binary systems compared with single stars \citep{Harris12, Akeson19, Barenfeld19}.  
\par
The role of enhanced planetesimal drift in tight binary systems has not been investigated in detail.  The effect was included in the multi-zone coagulation simulations done by \citet{Silsbee21}, using the same prescription for drift as in \citet{Rafikov15a}.  They found radial drift to have a non-zero, but not dominant effect on the planetesimal coagulation process.  However the simplified expression for the drift rate, which always results in inwards-directed drift, likely artificially lowered the importance of the effect. 
\par
In this work we revisit the calculation of the radial drift of a planetesimal in an eccentric gaseous disk.  In Section \ref{sect:exactCalc}, we use numerical integration to calculate the drift rate exactly as a function of planetesimal eccentricity given a power law model for the radial variation of gas density and eccentricity.  In Section \ref{sect:sustainedMigration}, we calculate also the effect of gas drag on the planetesimal's eccentricity vector.  Combining this with the secular equations for the evolution of planetesimal eccentricity, we identify two scenarios in which planetesimals undergo sustained outwards migration.  In Section \ref{sect:discuss}, we discuss the ramifications of these calculations, and some limitations and uncertainties.  In Section \ref{sect:pform}, we calculate the evolution of the planetesimal surface density in one disk model, showing dramatic enhancement in specific locations.  In Section \ref{sect:Adachi}, we quantitatively compare our results with the previously adopted approximation. In Sections \ref{sect:diskParams} - \ref{sect:diskThermo}, we discuss some uncertainties and limitations of our model.  In Section \ref{sect:conclusion}, we briefly summarize our conclusions.  Appendix \ref{append:simpleModel} develops a perturbative approximation that allows the main results to be understood analytically, but is not adequately accurate for quantitative calculations.
\par

\section{Calculation of radial drift in eccentric disks}
\label{sect:exactCalc}
In this section, we describe the equations used to numerically calculate the radial drift rate of planetesimals due to gas drag in an eccentric disk.  We consider the case of planetesimals with Stokes number much greater than 1, so that they may be assumed to follow Keplerian orbits. We limit ourselves to a disk with no warp, and apsidally aligned streamlines, and we assume the planetesimal to orbit in the disk midplane.  We take the eccentricity $e_d$ and surface density at pericenter $\Sigma_{\rm peri}$ and temperature at pericenter $T_{\rm peri}$ of the streamline with semi-major axis $a$ to be power law functions of $a$:
\begin{equation}
e_d(a) = e_0 \left(\frac{a_0}{a}\right)^q; \quad \quad \Sigma_{\rm peri}(a) = \Sigma_0 \left(\frac{a_0}{a}\right)^p; \quad \quad T_{\rm peri} = T_0 \left(\frac{a_0}{a}\right)^s.
\label{eq:eccentricityAndSurfaceDensity}
\end{equation}
We take a power-law model for disk eccentricity to allow semi-analytic computations, but we note that the true eccentricity structure is likely more complicated.  Simulations addressing the eccentricity of circumstellar disks in binary systems show a wide variety of behaviors.  We further discuss the role played by the disk eccentricity profile in Section \ref{sect:diskParams}. 
\par
\begin{figure*}[htp]
\centering
\includegraphics[width = \textwidth]{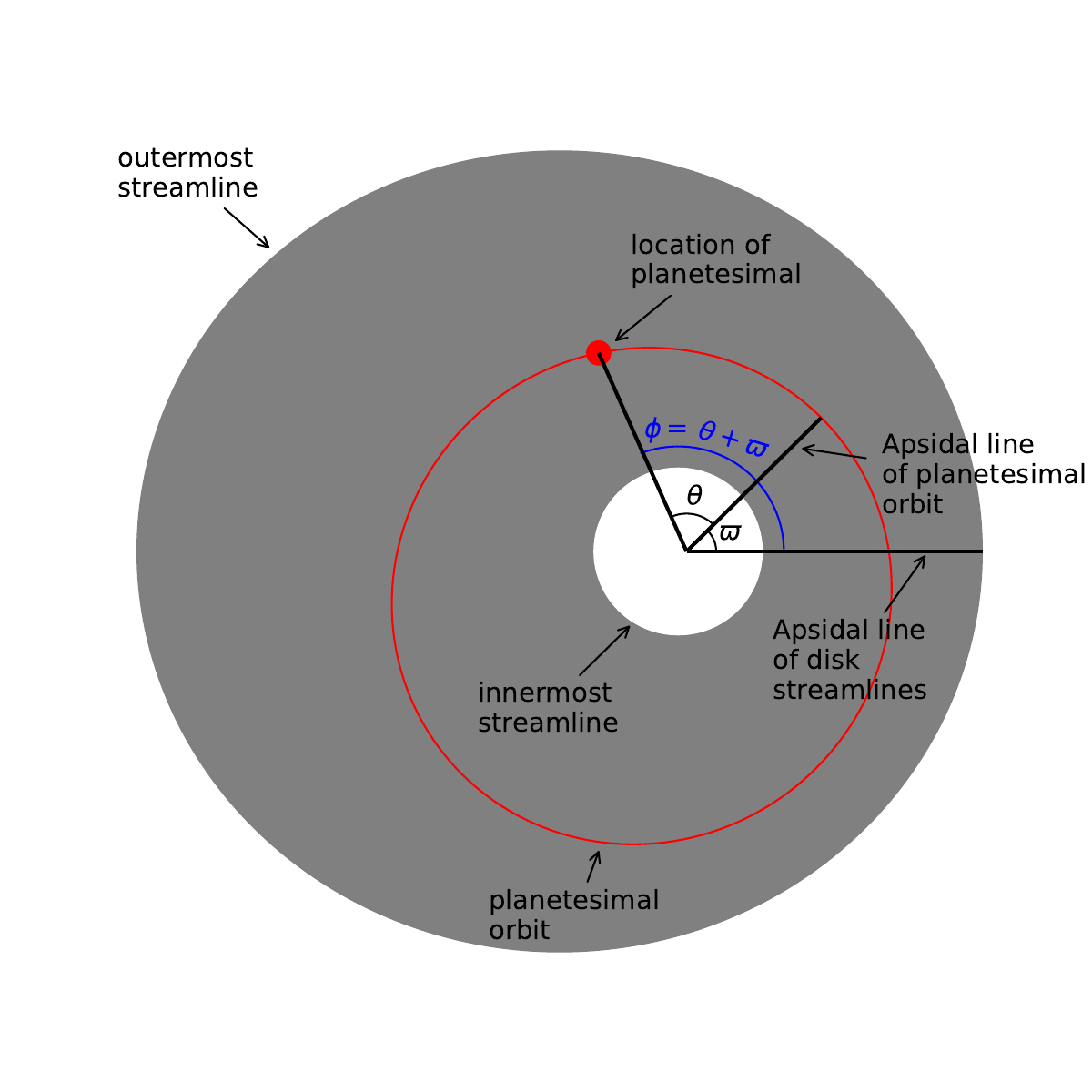}
\caption{Sketch of a planetesimal embedded in an eccentric disk with the angles used in this work labeled.  }
 \label{diskSketch}
\end{figure*}

 We consider the motion of a planetesimal orbiting in the disk midplane, with semi-major axis $a_p$, eccentricity $e_p$, and longitude of pericenter $\varpi_p$ with respect to the apsidal line of the disk.  Let $\theta$ denote the angle of the particle's radial vector with respect to the apsidal line of its orbit.  A sketch of the disk with the relevant angles labeled is shown in Figure \ref{diskSketch}.
\par
To calculate the planetesimal's velocity, we begin with the expression for the shape of a Keplerian orbit: 
\begin{equation}
r = \frac{a(1-e^2)}{1 + e \cos{\theta}}.
\label{eq:ellipseShape}
\end{equation}
We use conservation of energy and angular momentum to calculate $\dot r_p$ and $\dot \theta$.  That is to say, the following equations
\begin{equation}
\frac{1}{2} \dot r_p^2 + \frac{r_p^2 \dot \theta^2}{2} - \frac{GM}{r_p} = - \frac{GM}{2a_p}; \quad \quad \sqrt{GMa_p(1-e_p^2)} = r_p^2 \dot \theta,
\end{equation}
in conjunction with Equation \eqref{eq:ellipseShape}, result in these relations for the radial and azimuthal velocity:
\begin{equation}
v_p^r = \dot r_p = e_p \sin{\theta} \sqrt{\frac{GM}{a_p(1-e_p^2)}}; \quad \quad v_p^\theta = r_p \dot \theta = (1 + e_p \cos{\theta}) \sqrt{\frac{GM}{a_p(1-e_p^2)}}.
\label{eq:EccentricVelocities}
\end{equation}
Let $\phi = \theta + \varpi_p$ be the angle of the planetesimal's radial vector with respect to the apsidal line of the disk.  
Then, given the assumption that the gas streamlines are Keplerian ellipses, we substitute Equation \eqref{eq:eccentricityAndSurfaceDensity} for the gas eccentricity into Equation \eqref{eq:ellipseShape}, to yield an equation for the semi-major axis $a_g$ of the gas streamline at position $[r, \phi]$:
\begin{equation}
\frac{1 + e_0\tilde a^q \cos{\phi}}{1 - e_0^2 \tilde a^{2q}} \frac{r}{a_0} = 1/\tilde a,
\label{eq:rtoa}
\end{equation}
where $\tilde a = a_0/a_g$.  Throughout this work, we use the subscript ``$g$" to denote a property of the gas, and ``p" to denote a property of the planetesimal. We then solve Equation \eqref{eq:rtoa} numerically for $\tilde a$.  Once $a_g$ at the location of the planetesimal is known, the eccentricity $e_g$ is given by Equation \eqref{eq:eccentricityAndSurfaceDensity}.  In the absence of a pressure gradient, the radial and azimuthal velocities of the gas are given in terms of $a_g$ and $e_g$ by Equations \eqref{eq:EccentricVelocities}.  We then assume that the velocity is reduced by a factor of $\sqrt{1 - 2\eta(a_g)}$, independent of location along the streamline, where $\eta$ is given by 
\begin{equation}
\eta(a) = \frac{p + (s + 3)/2}{2} \times\left( \frac{c_s}{v_K(a)}\right)^2.
\label{eq:eta}
\end{equation}
For an axisymmetric disk, this is equivalent to Equation 3.4 from \citet{Adachi76}, noting that their $\alpha$ is equivalent to our $(p + 3/2 - s/2)$, their $\beta$ is our $s$, and the mean thermal velocity $c_m$ is related to the isothermal sound speed $c_s$ by $c_m = c_s\sqrt{8/\pi}$.
\par
Here $c_s$ is the isothermal sound speed, calculated assuming a mean mass per particle of 2.36 Daltons and a gas temperature appropriate for $r = a$ [see Equations \eqref{eq:eccentricityAndSurfaceDensity}, \eqref{eq:azimuthalVariations} and \eqref{eq:PRhoSigma}]. $v_K(a) = \sqrt{GM/a}$ is the Keplerian speed of a {\it circular} orbit with radius $a$.  $G$ is Newton's constant, and $M$ is the mass of the central star.  This results in a gas velocity given by 
\begin{equation}
v_g^r = e_g \sin{\phi} \sqrt{\frac{GM \left[1 - 2 \eta(a_g)\right]}{a_g(1-e_g^2)}}; \quad \quad v_g^\theta = (1 + e_g \cos{\phi}) \sqrt{\frac{GM \left[1 - 2 \eta(a_g)\right]}{a_g(1-e_g^2)}}.
\label{eq:GasVelocity}
\end{equation}

This treatment does not account for the azimuthal variation in velocity arising due to the pressure gradient, but as argued in Section \ref{sect:pressureEffects}, the effect of the azimuthal component of the pressure gradient on the gas velocity is small compared to that of the standard radial pressure gradient term present in an axisymmetric disk.  As the main aim of this paper is to elucidate the impact of disk eccentricity on the radial drift, and as the effect of pressure on the gas streamlines is uncertain, we leave further discussion of pressure effects to future work.
 \begin{figure*}[htp]
\centering
\includegraphics[width = \textwidth]{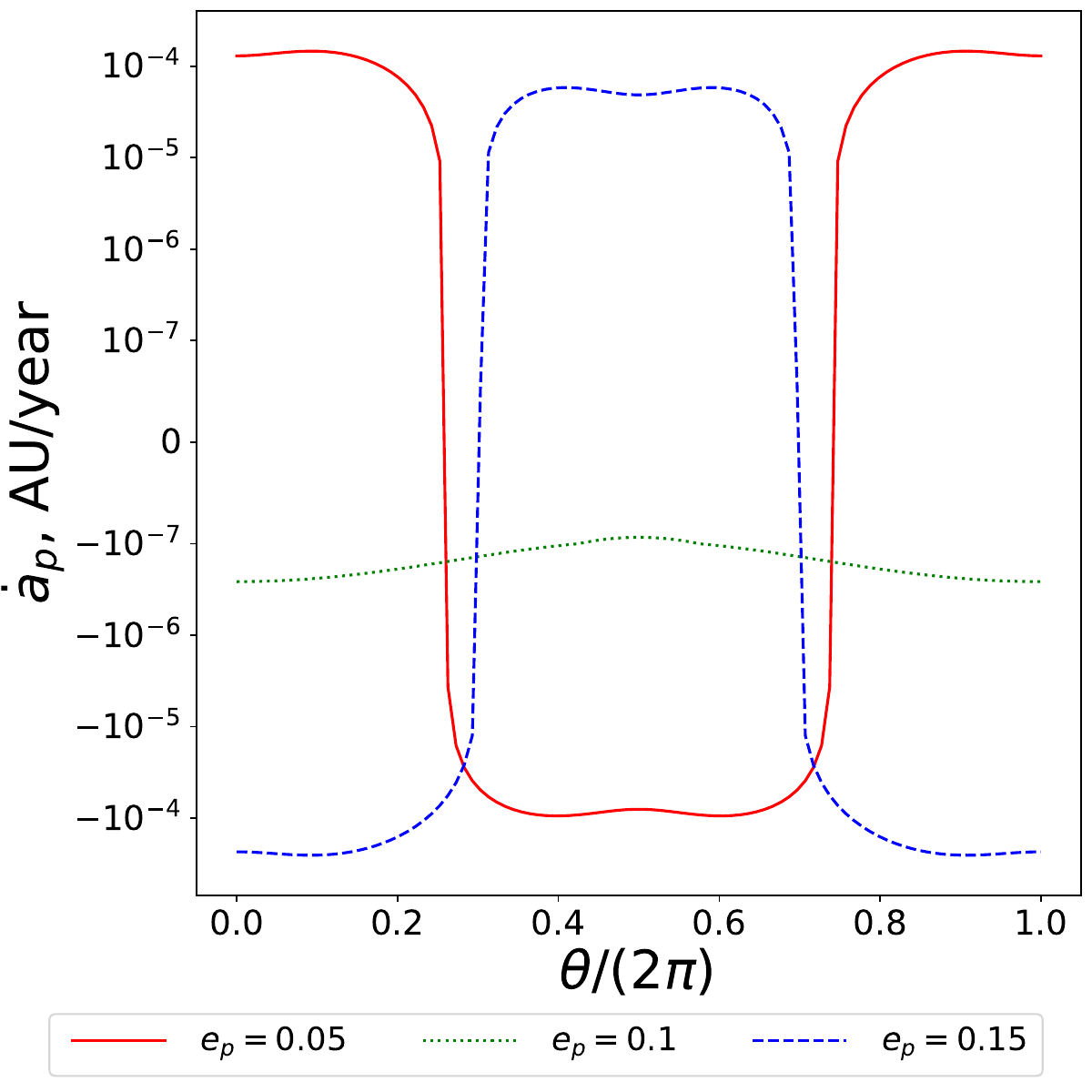}
\caption{Dependence of the instantaneous value of $\dot a_p$ on angle $\theta$ for a planetesimal apsidally aligned with the disk (i.e. with $\varpi_p = 0$.  Different color curves correspond to different values of the planetesimal eccentricity. 
 The disk is assumed to have the fiducial parameters listed in Table \ref{tab:fiducialParams}.   }
 \label{angular_dependence}
\end{figure*}
In order to calculate the drag force on the planetesimal, we need to know the gas density as a function of position.  The surface density distribution of a disk described by Equation \eqref{eq:eccentricityAndSurfaceDensity} is given by \citep[see][]{Statler99, Silsbee15}
\begin{equation}
\Sigma(a_g, \phi)=  \Sigma_{\rm peri}(a_g) \frac{1-e_g^2 +q e_g(1+e_g)}{1 - e_g^2 +q e_g(e_g + \cos{E})},
\label{eq:finalSigma}
\end{equation}
where $E$ is the eccentric anomaly at azimuthal angle $\phi$ of the streamline with semi-major axis $a_g$, and $\Sigma(a_g, \phi)$ is the surface density at azimuthal angle $\phi$ along the streamline with semi-major axis $a_g$.  $\Sigma_{\rm peri}(a_g) \equiv \Sigma(a_g, 0)$ is the surface density at pericenter.  Then, assuming assuming the disk to be in vertical hydrostatic equilibrium and vertically isothermal, the mid-plane density along the apsidal line of the disk is given by 
\begin{equation}
\rho_{\rm peri}(a_g) = \frac{\Sigma_{\rm peri}}{\sqrt{2\pi}H(r_{\rm peri})},
\label{eq:pericenterDensity}
\end{equation}
where 
$H(r) = rc_s/v_K(r)$. 
\par

We assume an adiabatic equation of state on an orbital timescale, and that all parts of the disk are in vertical hydrostatic equilibrium.  In Section \ref{sect:diskThermo}, we explore the effect of other treatments of the disk thermodynamics and vertical relaxation timescale.
\par
In order to calculate the variation of mid-plane density $\rho_g$ with $E$, we use the following equations.  At eccentric anomaly $E$, we let 
 \begin{equation}
 P = A P_{\rm peri}; \quad \rho_g = B \rho_{\rm peri}; \quad \Sigma = C \Sigma_{\rm peri}; \quad r = D r_{\rm peri}; \quad H = F H_{\rm peri}; \quad T = \chi T_{\rm peri}.
 \label{eq:azimuthalVariations}
 \end{equation}
 $C$ can be found easily from Equation \eqref{eq:finalSigma}, and $D$ from the relation $r = a(1-e\cos{E})$.  In order to solve for $A$, we need 4 other equations.  These are:
 \begin{enumerate}
 \item Equation of State: $P/P_{\rm peri} = (\rho/\rho_{\rm peri})^\gamma \rightarrow A = B^\gamma$
 \item Ideal Gas Law: $P \propto \rho_g T \rightarrow A = B\chi$
 \item $H = c_sr/v_K(r) \propto \sqrt{T}r^{3/2} \rightarrow F = \chi^{1/2}D^{3/2}$
 \item $\Sigma \sim H \rho \rightarrow C = FB$
 \end{enumerate}
 Here $\gamma$ is the adiabatic index.  We take this to be 10/7, appropriate for a mixture with a $5:1$ number density ratio of H$_2$ and He, assuming 5 accessible degrees of freedom per H$_2$ molecule.  The adiabatic index of H$_2$ has some temperature dependence, due to the change in number of accessible degrees of freedom; see Figure 1 of \citet{Sharda19}.
 Solution of these equations leads to the relations
 \begin{equation}
 \frac{\rho_g}{\rho_{\rm peri}} = \left(\frac{C^2}{D^3}\right)^\frac{1}{\gamma +1}; \quad \quad \frac{P}{P_{\rm peri}} = \left(\frac{C^2}{D^3}\right)^\frac{\gamma}{\gamma +1}; \quad \quad \frac{T}{T_{\rm peri}} = \left(\frac{C^2}{D^3}\right)^\frac{\gamma-1}{\gamma +1}.
 \label{eq:PRhoSigma}
 \end{equation}
 At azimuthal angle $\phi$, we take the mid-plane gas density $\rho_g$ given by Equations \eqref{eq:pericenterDensity} and \eqref{eq:PRhoSigma}.
 \par  
 The instantaneous drag force on a planetesimal of radius $R$ is given by 
\begin{equation}
\vec F_d = - \frac{\pi C_d R^2 \rho_g v_{\rm rel} |v_{\rm rel}|}{2},
\label{eq:dragForce}
\end{equation}
where $C_d$ is a constant of order unity that we take to be 0.5 in our numerical calculations.
\par
We then calculate the rate of change in semi-major axis using the Gauss planetary equation relating the force on a body to the change in semi-major axis; see Equation 1.200 from \citet{Tremaine23}:
\begin{equation}
\dot a_p = \frac{2a_p^{3/2} e_p F_r \sin{\theta}}{m_p\sqrt{GM(1 - e_p^2)}}  + \frac{2a_p^{3/2} F_\theta (1+e_p \cos{\theta})}{m_p \sqrt{GM(1-e_p^2)}},
\label{eq:almostFinalInspiralEquation}
\end{equation}
where  $m_p = 4\pi \rho_p R^3/3$ is the mass of the planetesimal.  In our calculations, we assume the planetesimal density $\rho_p = 2$ g cm$^{-3}$. 
\par
Figure \ref{angular_dependence} shows the instantaneous value of $\dot a_p$ as a function of $\theta$ for different values of the planetesimal eccentricity.  We assume that the planetesimal is apsidally aligned with the disk (i.e. $\varpi_p = 0$), and the disk parameters are those given in Table \ref{tab:fiducialParams}.  In order to calculate the orbit-averaged inspiral rate, we average the result in Equation \eqref{eq:almostFinalInspiralEquation} over $\theta$, weighted by the inverse angular velocity, to get
\begin{equation}
\langle \dot a_p(a_p, e_p) \rangle = \frac{(1 - e_p^2)^{3/2}}{2 \pi } \int_0^{2\pi} \frac{\dot a_p(a_p, e_p, \theta)}{(1 + e_p\cos{\theta})^2} d \theta.
\label{eq:finalInspiralEquation}
\end{equation}

\begin{table}[h!]
  \begin{center}
    \caption{{\bf List of symbols}}
    \label{tab:fiducialParams}
  \begin{tabular}{l l l }
        symbol & meaning & fiducial value \\
        \hline
     $a_0$ & reference value of the semi-major axis & 1 AU\\
     $\Sigma_0$ & surface density at $a_0$ & 1000 g/cm$^2$ \\
     $e_0$ & disk eccentricity at $a_0$ & 0.1\\
     $T_0$ & pericenter gas temperature at $a_0$ & 200 K\\
     $p$ & exponent of $\Sigma_p$ dependence on $a^{-1}$ & 1 \\
     $q$ & exponent of $e_g$ dependence on $a^{-1}$ & -1\\
     $s$ & exponent of $T_p$ dependence on $a^{-1}$ & 1/2\\
     $M$ & mass of central star & 1 $M_\odot$\\
     $R$ & planetesimal radius & 1 km\\
     $\rho_p$ & planetesimal density & 2 g cm$^{-3}$\\
     $\gamma$ & adiabatic index of the gas & 10/7 \\
     $\nu$ & mass ratio of secondary to primary & 1 \\
     $a_b$ & semi-major axis of binary orbit & 20 AU \\
     $e_b$ & eccentricity of binary orbit & 0.4\\
     $a_{\rm in}$ & semi-major axis of innermost streamline & 0.1 AU\\
     $a_{\rm out}$ & semi-major axis of outermost streamline & 5 AU\\
     $C_d$ & drag coefficient & 0.5 \\
     $a_p$ & semi-major axis of planetesimal's orbit & - \\
     $e_p$ & scalar planetesimal eccentricity & - \\
     $\theta$ & angle of planetesimal w.r.t. its apsidal line & - \\
     $\phi$ & angle of planeteismal w.r.t. the disk's apsidal line & - \\
     $\varpi_p$ & apsidal angle of planetesimal w.r.t. the disk & - \\
     $k_p$ & $e_p \cos{\varpi_p}$ & - \\
     $h_p$ & $e_p \sin{\varpi_p}$ & -

    \end{tabular}
  \end{center}
\vspace{-.5cm}
\end{table}

 \begin{figure*}[htp]
\centering
\includegraphics[width = \textwidth]{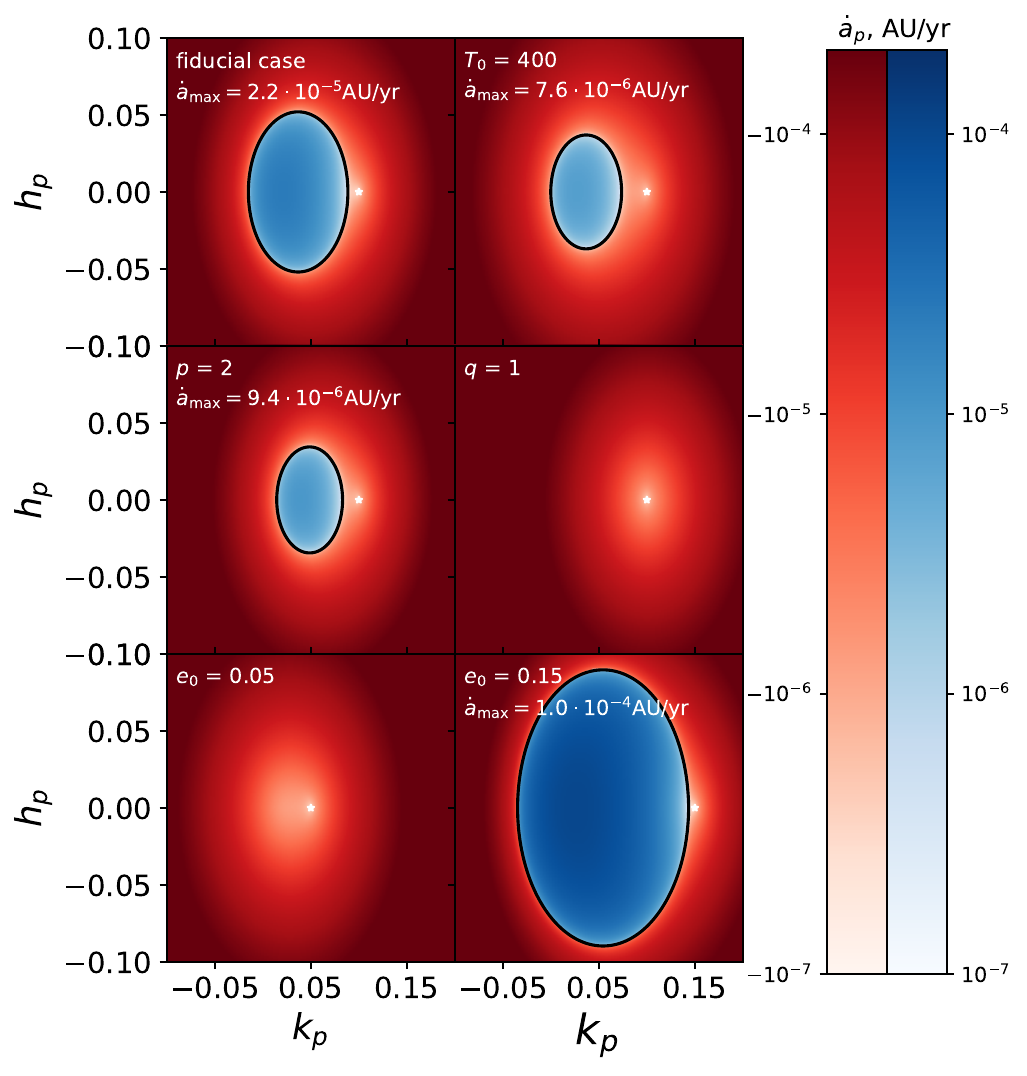}
\caption{Inspiral rate for a 1-km planetesimal with semi-major axis 1 AU.  This is plotted in the space of the components $k_p$ and $h_p$ of the planetesimal's eccentricity vector [see Equation \eqref{eq:kandh}].  Negative values of the drift, corresponding to motion towards the central star, are shown in red.  Positive values, corresponding to drift away from the star, are shown in blue.  The top left panel corresponds to a disk with the fiducial parameters as shown in Table \ref{tab:fiducialParams}.  In each other panel, one parameter is varied, as labeled.  The disk eccentricity vector at the planetesimal semi-major axis is shown by the white star.  The color scale does not extend beyond $\pm 2 \times 10^{-4}$ AU/yr, leading to a saturation of the red colors.  The blue regions are unaffected by this.  For the panels containing a region with an outward-directed drift, the maximum value of the drift rate is labeled. }
 \label{inspiralMap}
\end{figure*}
Figure \ref{inspiralMap} shows the inspiral rate of a 1-km planetesimal in a variety of model disks in the space of the two components $k_p$ and $h_p$ of the eccentricity vector of a planetesimal embedded in the disk.  These are defined in terms of $e_p$ and $\varpi_p$ as 
\begin{equation}
k_p = e_p \cos{\varpi_p}; \quad \quad h_p = e_p \sin{\varpi_p}.
\label{eq:kandh}
\end{equation}
A planetesimal orbit apsidally aligned with the disk lies along the positive $k_p$ axis.  Negative values of the drift rate, corresponding to drift towards the central star are shown in red.  Positive values, corresponding to outwards planetesimal migration are shown in blue.
\par
 The top left panel shows the results for a disk and planetesimal with the fiducial parameters in Table \ref{tab:fiducialParams}. In this case, the region of parameter space, corresponding to outwards migration is centered on the $k_p$ axis, at a value of $k_p$ roughly 1/2 the local disk eccentricity. 
 \par
 The cause of outwards migration can be understood qualitatively as follows.  For an apsidally aligned particle with eccentricity less than the local disk eccentricity, at pericenter the particle is interacting with a streamline with larger semi-major axis.  Therefore, ignoring the effect of pressure on gas velocity, i.e. setting $\eta$ to zero in Equation \eqref{eq:GasVelocity}, the gas velocity is higher than the particle velocity at pericenter, since larger semi-major axis corresponds to larger (less negative) specific energy.  The opposite is true at apocenter.  Therefore the particle gains energy at pericenter and loses it at apocenter.  This behavior can be seen in the red curve in Figure \ref{angular_dependence}.  Note that this is the reverse of how it works in an axisymmetric disk, where the planetesimal gains energy at apocenter and loses it at pericenter.  In an eccentric disk, the energy gain at pericenter can be larger than the corresponding loss as apocenter for two reasons.  First the gas density is generally higher at pericenter than at apocenter.  Second, even including a non-zero $\eta$, when the disk eccentricity gradient is sufficiently steep, the relative velocity at pericenter is larger than that at apocenter for moderate values of the eccentricity difference [see Equations \eqref{eq:vrelShort} - \eqref{eq:ca}].  
 \par
 Apsidally aligned planetesimals with eccentricity larger than the local disk eccentricity gain energy at apocenter and lose it at pericenter (see the blue curve of Figure \ref{angular_dependence}).  However, in this case the energy losses at pericenter are greater than the gains at apocenter, so the net drift is inwards.  A particle with eccentricity exactly equal to the local disk eccentricity (Figure \ref{angular_dependence}, green curve) experiences motion relative to the gas only because of the pressure gradient, and therefore loses energy over its entire orbit.
 \par
 In the top right panel, the disk temperature is increased by a factor of two relative to the fiducial value, i.e. $T_0 = 400$ K.  This has two effects.  First, since surface density is held fixed, it decreases the mid-plane gas density, thus slightly slowing the radial drift.  Second, it increases the value of $\eta$, [see Equation \eqref{eq:eta}], thus reducing the amount of area in the figure corresponding to outwards motion.
\par
The middle left panel shows the case with $p = 2$, i.e. a steeper radial decrease of surface density.  In the blue region of the plot, the local gas streamline at planetesimal pericenter has higher semi-major axis than the local gas streamline at planetesimal apocenter.  Therefore, counterintuitively, a steeper radial decrease of surface density increases the ratio of the drag force at apocenter relative to that at pericenter.  As discussed above, for orbits with eccentricity vector lying in the blue-shaded region, the planetesimal is {\it gaining} energy at pericenter, whereas at apocenter it is losing energy.  Therefore, a steeper surface density profile decreases the area in which the planetesimals drift outwards.
\par
The middle right panel shows the case in which $q = 1$, i.e. the magnitude of the disk eccentricity increases inwards, rather than outwards.  In this case, along the $k_p$ axis, for values of $k_p$ less than the disk eccentricity, the inspiral rate is faster than the standard case with $q = -1$.  This is because the headwind felt at apocenter increases faster with the quantity $\epsilon = e_0 - k_p$, than does the tailwind at pericenter [see Equations \eqref{eq:cp} and \eqref{eq:ca}].  Along the $k_p$ axis, for values of $k_p$ greater than the disk eccentricity, the inspiral rate is slower than in the fiducial case, but still negative.  Ignoring the effect of the pressure gradient in the disk, planetesimals with $h_p = 0$ and $k_0 > e_0$ feel a headwind at pericenter, and a tailwind at apocenter for $q = 1$.  While the tailwind at apocenter is stronger than the headwind at pericenter [see Equations \eqref{eq:cp} and \eqref{eq:ca}], the drift is directed inwards due to a combination of the higher gas density at pericenter and the effect of the pressure gradient term.
\par
The bottom left panel shows a case in which disk streamlines with a given semi-major axis have a factor of two lower eccentricity.  In this case, the effect of the pressure gradient on the radial velocity becomes more important than the disk eccentricity, and therefore the drift is always inwards.  In the bottom right panel, we have increased the disk eccentricity, and the region of outward radial drift is correspondingly larger.

\section{Sustained outwards drift}
\label{sect:sustainedMigration}
In the previous section we ignored the effect of gas drag on the eccentricity vector.  When only gas drag is acting on the planetesimal, the planetesimal's eccentricity vector will become aligned with that of the local streamline, at which point drift is determined solely by the pressure gradient.  In this case, in the absence of a local pressure maximum, outward migration will only be a transient phenomenon resulting from a non-equilibrium initial condition.  However, as we show in this section, sustained outwards drift is possible in the case that the planetesimal eccentricity is perturbed by something in addition to gas drag.  In this section, we calculate the evolution of the planetesimal eccentricity vector and semi-major axis under the combined influence of gas drag and secular gravitational perturbations from the disk and a binary companion star.
\par
To calculate the evolution of $k_p$ and $h_p$ from the gas drag, we insert our expression for the drag force given by Equation \eqref{eq:dragForce} into the Gauss equations for the evolution of eccentricity and apsidal angle [see Equations 1.200 from \citet{Tremaine23}].
\begin{equation}
\dot e_p = \sqrt{\frac{a_p(1-e_p^2)}{GM}} \left[\frac{F_r \sin{\theta}}{m_p} + \frac{F_\phi}{m_p}(\cos{E} + \cos{\theta})\right];
\label{eq:gaussedot}
\end{equation}
\begin{equation}
\dot \varpi_p = \sqrt{\frac{a_p(1-e_p^2)}{GMe^2}} \left[\frac{-F_r \cos{\theta}}{m_p} + \frac{\sin{\theta}(2 + e_p \cos{\theta})F_\phi}{(1 + e_p \cos{\theta})m_p}\right].
\label{eq:gaussvarpidot}
\end{equation}

We then average these over $\theta$ as done in Equation \eqref{eq:finalInspiralEquation}.  $\dot k_p$ and $\dot h_p$ are given by 
\begin{equation}
\dot k_p = \dot e_p \cos{\varpi_p} - e_p \sin{\varpi_p} \dot \varpi_p; \quad \quad \dot h_p = \dot e_p \sin{\varpi_p} + e_p \cos{\varpi_p} \dot \varpi_p.
\label{eq:khdot}
\end{equation} 
In addition to gas drag, we assume that the planetesimal is perturbed by an external companion star with mass ratio $\nu$, which orbits with semi-major axis $a_b$, and orbital eccentricity $e_b$.  The orbit of the companion star is assumed to be co-planar and apsidally aligned with the gas disk.  We consider only the secular perturbations from the companion star on the planetesimal orbit.  With this set-up, we calculate the evolution of $k_p$ and $h_p$ using Equations 3-8 from \citet{Rafikov15b}, but calculating $\dot k_p^{\rm drag}$ and $\dot h_p^{\rm drag}$ as in the current paper.  That is to say, we let 
\begin{equation}
\dot k_p = -Ah_p + \dot k_p^{\rm drag}; \quad \quad \dot h_p = Ak_p + B_b + B_d + \dot h_p^{\rm drag},
\end{equation}
where $A$ is the precession frequency of the free eccentricity, given by the sum of terms corresponding to the disk and the binary
\begin{equation}
    A = A_d + A_b = 2\pi \frac{G \Sigma_p(a_p)}{n_p a_p} \psi_1 + \frac{3}{4} n_p \nu \left(\frac{a_p}{a_b}\right)^3,
\end{equation}
and the eccentricity excitation terms $B_d$ and $B_b$ are given by 
\begin{equation}
    B_d = \pi \frac{G \Sigma_p(a_p)}{n_p a_p} e_d(a_p) \psi_2 \quad \quad B_b = -\frac{15}{16} n_p \nu \left(\frac{a_p}{a_b}\right)^4e_b.
\end{equation}
$n_p$ is the mean motion of the particle.  $\dot k_p^{\rm drag}$ and $\dot h_p^{\rm drag}$ are calculated numerically using Equations \ref{eq:gaussedot} - \ref{eq:khdot} from the current paper, rather than the approximation in \citet{Rafikov15b}.
The coefficients $\psi_1$ and $\psi_2$, which take into account the location of the disk edges, are given in Equations A.33 and A.34 of \citet{Silsbee15}. 
 \begin{figure*}[htp]
\centering
\includegraphics[width = \textwidth]{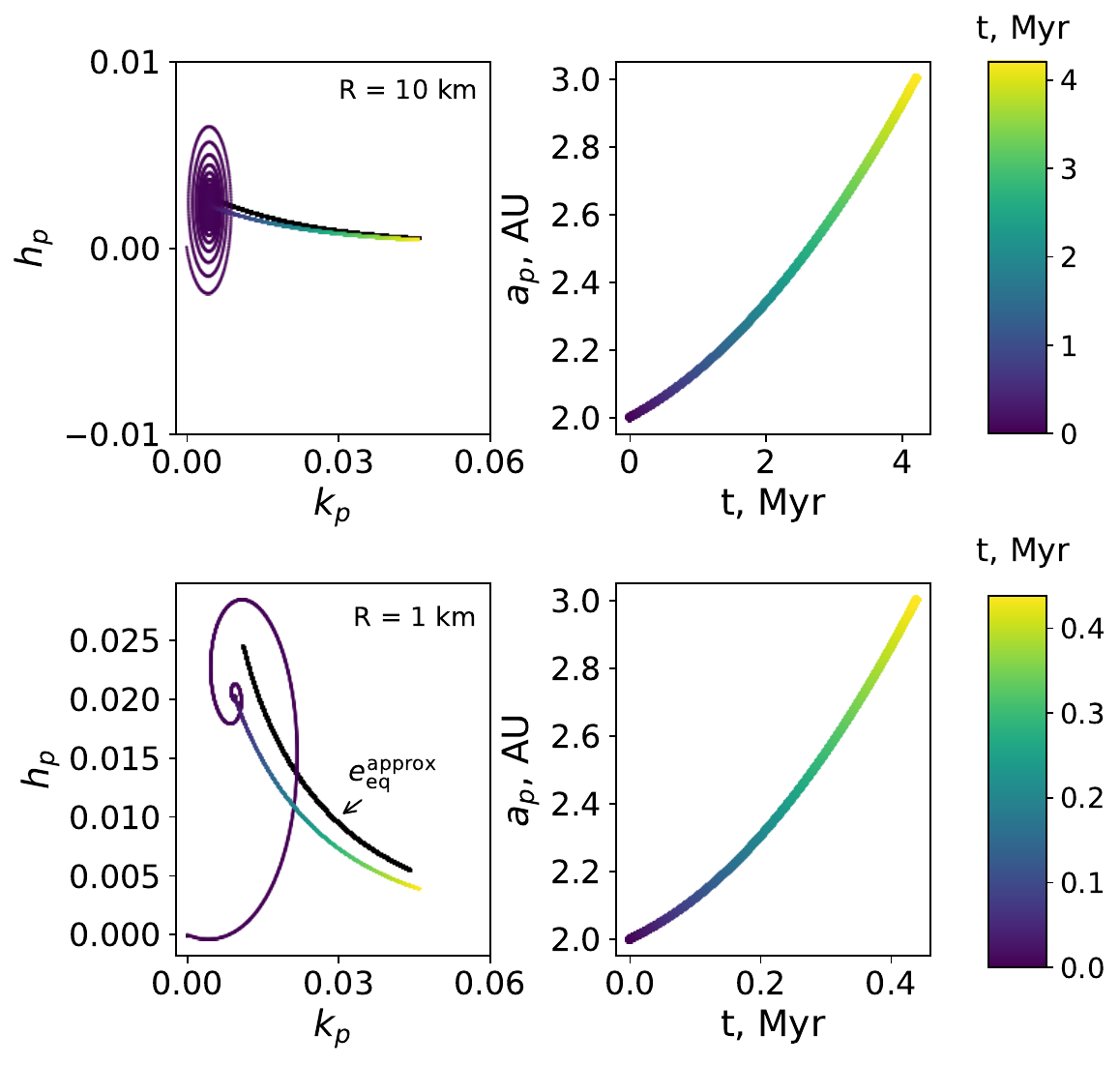}
\caption{Evolution of planetesimal eccentricity vector (left panels) and semi-major axis (right panels) under the combined influence of gas drag, and secular gravitational perturbations from the disk and binary companion.  The color corresponds to the amount of time since the beginning of the simulation, as shown in the color-bars on the right.  The top panels show the trajectory of a 10 km planetesimal, and the bottom a 1 km planetesimal.  This assumes the fiducial system parameters [see Table \ref{tab:fiducialParams}], except $\Sigma_0 = 300$ g cm$^{-2}$, and $e_0 = 0.05$.  The black line (almost overlapping the colored line in the upper panel) shows the equilibrium eccentricity calculated using Equations 22 and 23 from \citet{Rafikov15b}.}
 \label{withPerturber}
\end{figure*}
\par
We identify two sets of parameters leading to a region of the disk in which sustained outwards drift is possible.  One case is that of a low-mass disk apsidally aligned with the binary orbit, in which gravitational perturbations are dominated by the binary.  The second set of parameters corresponds to a high-mass disk in which gravitational perturbations from the disk dominate over those from the binary companion.
\subsection{Sustained outwards drift in a low-mass disk}
Figure \ref{withPerturber} shows the results of two simulations of the orbital evolution of planetesimals in a low-mass disk ($\Sigma_0 = 300$ g cm$^{-2}$), subject to gravitational perturbations from a binary companion.  The left panels show the evolution of the planetesimal's eccentricity vector.  The right panels show the evolution of the semi-major axis.  Time is given by the color scale (see color-bars on the right).  The top panels correspond to a planetesimal with radius 10 km, and the bottom panels to one with radius 1 km.  The time required to drift from 2 to 3 AU is about a factor of 10 higher for the 10 km planetesimal than for the 1 km one.
\par
In the left panels, the black line shows the equilibrium eccentricity calculated with Equations 22 and 23 from \citet{Rafikov15b}\footnote{We note that in deriving Equation 14 in \citet{Rafikov15b}, they implicitly assumed the mid-plane gas density to be $\Sigma_g/H$, rather than $\Sigma_g/(H\sqrt{2\pi})$ as in the present work.  In creating the black line on Figure \ref{withPerturber}, we multiplied ``h" in Equations 14 and 31 of \citet{Rafikov15b} by $\sqrt{2\pi}$ to account for this difference.}.  This is visibly, though not dramatically different from the colored curve in the case with $R = 1$ km, and lies very close to the colored line for $R = 10$ km.  This shows that the approximate calculation in \citet{Rafikov15b, Rafikov15a} adequately describes the planetesimals' eccentricity evolution.  We also calculated the true equilibrium eccentricity as a function of $a_p$ by evolving $k_p$ and $h_p$ to steady state while holding $a_p$ fixed.  We find that for $a_p \geq 2.02$ AU, $h_p$ and $k_p$ given by the colored curve deviate from their equilibrium values by less than $10^{-4}$.  In other words, after a brief transient behavior (shown by the spiral portion of the colored curve), at all times the eccentricity vector remains very near the equilibrium value at the current semi-major axis. This is a consequence of the fact that the timescale for evolution of the semi-major axis is much longer than the timescale for gas drag to equilibrate the eccentricity.
\subsection{Sustained outwards drift in a high-mass disk}
\label{sect:highMass}
We consider also a case in which the disk is massive enough (for the chosen binary separation and mass ratio) to have a dominant effect on the gravitational perturbations acting on the planetesimals.  In this case, for our fiducial values of $p$ and $q$, the equilibrium planetesimal eccentricity is larger than the disk eccentricity, and therefore does not support sustained outwards drift. 
\par
\begin{figure*}[htp]
\centering
\includegraphics[width = \textwidth]{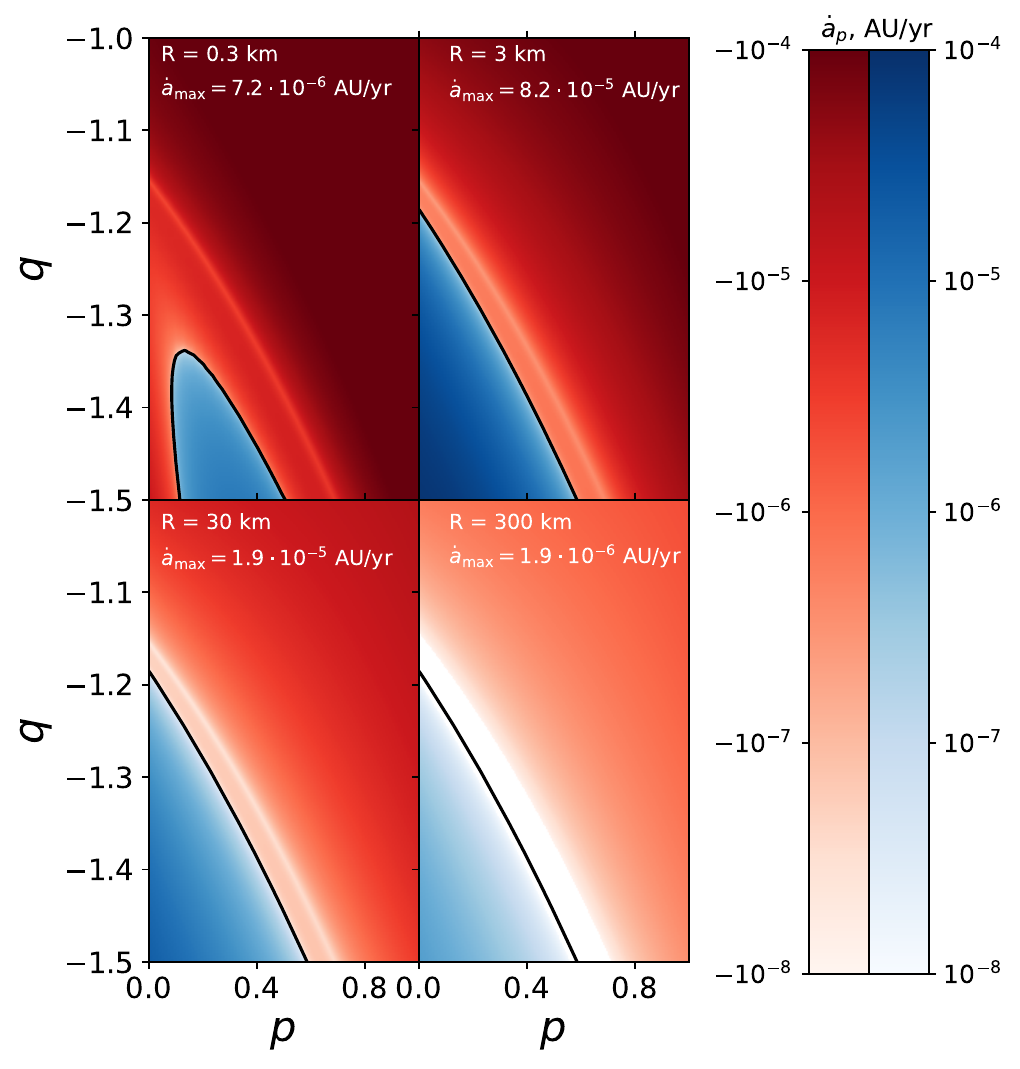}
\caption{Radial drift rate at 1 AU assuming the planetesimal's eccentricity vector to be at its equilibrium value.  This is in a system with the fiducial parameters in Table \ref{tab:fiducialParams}, except that $\Sigma_0 = 10,\!000$ g cm$^{-2}$, $a_{\rm out} = 2.5$ AU, and $p$ and $q$ are labeled on the axes.  The planetesimal size is labeled on each panel, and the maximum outwards drift rate is labeled on each panel.  The color scale does not extend beyond $10^{-4}$, which leads to a saturation of the red color in the top two panels.  
}
 \label{denseInspiral}
\end{figure*}

However, there are values of $p$ and $q$ for which sustained outwards drift is possible.  Figure \ref{denseInspiral} shows the drift rate at 1 AU for a system with the fiducial parameters, except $\Sigma_0 = 10,\!000$ g cm$^{-2}$, and $p$ and $q$ vary as shown on the axes.  In this case we kept the inner edge of the disk at $a_{\rm in} = 0.1$ AU, but placed the outer edge of the disk at $a_{\rm out} = 2.5$ AU.  This avoids streamline crossing, which would occur for models with $q < -1$ if we allowed the disk to extend to 5 AU. 
\par
The maximum value of the (outward-directed) radial drift rate is labeled on each panel.  In this case we see that substantial drift within the disk lifetime is possible even for $\sim \!100$ km planetesimals.  Since we assume that gravitational perturbations from the disk dominate over those from the binary, both the perturbing gravitational force and the force of gas drag are directly proportional to the disk mass.  As can be shown rigorously from the equations in \citet{Rafikov15b}, the equilibrium planetesimal eccentricity becomes independent of disk mass in this case.  Therefore, the inspiral rate for planetesimals of a given size is directly proportional to the disk mass, since the drag force for a planetesimal on a particular orbit is directly proportional to the gas density.  Above a certain size, the planetesimal eccentricity vectors become aligned \citep{Rafikov15b}, and therefore the drift rate is inversely proportional to planetesimal size.  Thus, the peak drift rate for 300 km planetesimals is almost exactly a factor of 10 slower than that for 30 km planetesimals.

\section{Discussion}
\label{sect:discuss}
\subsection{Impact on planet formation}
\label{sect:pform}
In the previous section, we pointed out two scenarios in which outward migration of planetesimals could occur.  Since the radial pressure gradient yields a bias for inwards drift, it is difficult to construct a scenario in which planetesimal drift is outwards throughout the entire disk.  Thus outward migration, if present, will be limited to a specific part of the disk.  For a given planetesimal size, we denote the boundaries of such a region $a_1$ and $a_2$, with $a_2 > a_1$.  The point $a_2$ is a stable attractor, in the sense that all planetesimals in the neighborhood of $a_2$ drift towards $a_2$.  In contrast, $a_1$ is an unstable stationary point in the sense that in the neighborhood around $a_1$, all planetesimals (except those with $a$ exactly equal $a_1$) drift away from $a_1$.  Over time, planetesimals become concentrated around $a_2$, and if the location of $a_2$ is almost independent of planetesimal size over a wide range of sizes, the overall surface density of planetesimals may become strongly enhanced at $a_2$.  The location of the region of outward migration (and therefore the location of the points $a_1$ and $a_2$) depends on the choice of disk parameters.  We discuss some of these dependencies in Section \ref{sect:diskParams}, however because of the complicated dependence of the disk gravity on parameters, and because the sign of the drift is determined by multiple effects acting in different directions, we are not able to give, even in the context of the current model a concise analytic expression for the location of $a_1$ and $a_2$.
\par
 Possible concentration of planetesimals due to radial drift was discussed briefly in \citet{Rafikov15a}, however it was not found to be a dramatic effect.  The authors of that work used a simplified model for the drift, taking Equation 4.21 from \citet{Adachi76} (derived and intended to be used for an axisymmetric disk) and replacing the planetesimal eccentricity with the relative planetesimal gas eccentricity.  Unlike the current treatment, this results in a drift rate that is always inwards, limited from below by the rate arising from the pressure gradient.  Roughly speaking in such a model, the peak planetesimal density enhancement in a local region of slower drift is equal to the ratio of the drift rate in the surrounding region to that in the local region (assuming that the region of slow drift is small enough that the planetesimal flux is approximately constant across it).  In contrast, in the current model which allows for convergent planetesimal migration, given sufficient time, the density contrast can grow to be arbitrarily large.
\par
We simulated the evolution of the planetesimal surface density under the action of radial drift.  Motivated by the results in Figure 4, we consider a system in which $p = 0.3$, $q = -1.4$, $\Sigma_0 = 10,\!000$ g cm$^{-2}$, $a_{\rm in} = 0.1$ AU, and $a_{\rm out} = 2.5$ AU, and the rest of the parameters take their fiducial values.  We began with a population of planetesimals distributed smoothly throughout the disk with semi-major axes from 0.2 to 2 AU assuming a local planetesimal to gas mass ratio of 0.2\%.  We calculated the drift rate assuming planetesimals to orbit at their equilibrium eccentricity, and then used that drift rate to solve for the evolution of the surface density distribution.  
\par
\begin{figure*}[htp]
\centering
\includegraphics[width = \textwidth]{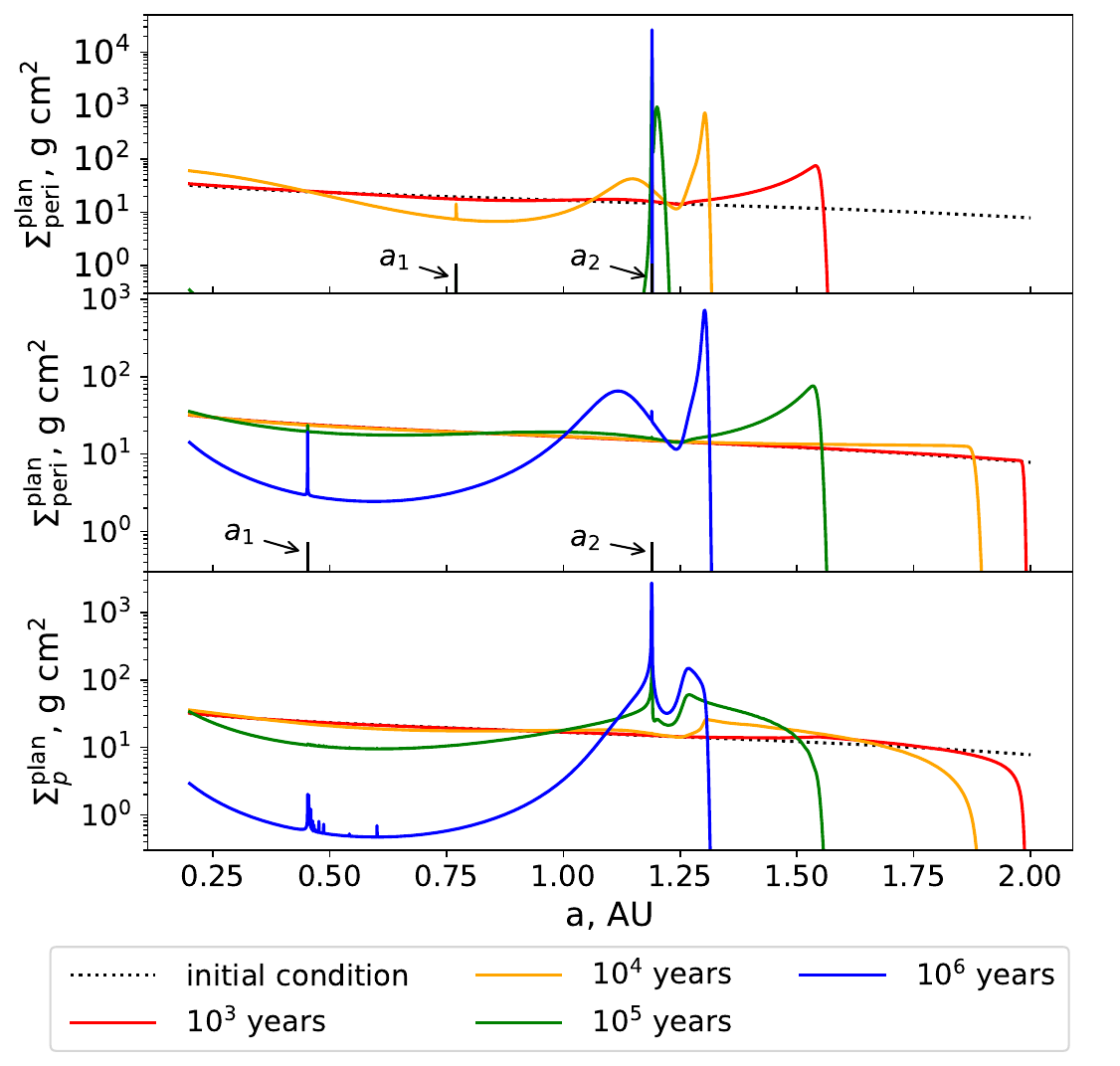}
\caption{Evolution of the planetesimal surface density under the effect of radial drift: each curve shows the surface density of planetesimals along the apsidal line of the disk as a function of semi-major axis at a different time since the beginning of the simulation, as labeled in the legend.  The disk and binary properties are identical to that of the ``high mass" case discussed in Section \ref{sect:highMass}.  The top panel corresponds to 1 km planetesimals, the middle panel to 100 km planetesimals, and the bottom panel to a size distribution between 1 and 100 km proportional to $R^{-7/2}$.  In the top and middle panels we have labeled the values $a_{1}$ and $a_{2}$ corresponding to the endpoints of the range of outward migration for planetesimals of that size. 
}
 \label{surfaceDensity}
\end{figure*}

Figure \ref{surfaceDensity} shows the planetesimal surface density at pericenter as a function of semi-major axis for different times between 0 and $10^6$ years.  The top panel shows the case in which all planetesimals are one km in size.  These planetesimals experience outwards drift between $a_1 = 0.77$ AU and $a_2 = 1.19$ AU (labeled on the panel).  This results in a few noticeable features on the graph.  First, a single-pixel spike develops at $a_1$ after 1000 years.  This is related to the discretization of semi-major axis space, and the fact that the drift rate drops to zero at $a_1$.  Since this spike is just composed of the planetesimals that were initially at that location, it is no higher than the initial surface density at that location.   
\par
In the vicinity of $a_2$, we see two peaks growing and drifting towards $a_2$.  One peak just outside of $a_2$ is composed of all the particles in the outer disk that have drifted inwards, and one just inside $a_2$, is composed of planetesimals initially between $a_1$ and $a_2$.  By 1 Myr, they have merged together forming a single peak with width equal to the smallest resolvable width in the simulation, and height over 1000 times the original surface density at that point.  In principle if we considered infinitely narrow bins in semi-major axis space and let planetesimals drift indefinitely in a static disk with no self-interaction, the two spikes would become arbitrarily high but never merge. 
 However at high planetesimal density and small scales, other physics that we don't include (e.g. collisions and planetesimal-planetesimal scattering) play a role as well, thus altering the size distribution and smoothing the surface density distribution.  We include this plot to demonstrate that a strong enhancement of the surface density may occur, not to claim an accurate calculation of the profile of the surface density peak.
\par
The middle panel shows the case in which all planetesimals have a size of 100 km.  These experience outward drift between $a_1 = 0.45$ AU and $a_2 =1.19$ AU.  The same features are visible as in the top panel, but because the larger planetesimals drift more slowly, the surface density evolves more slowly and the two peaks around $a_2$ are still clearly distinguishable even after $10^6$ years.  The complicated shape of the peaks reflects the behavior of the drift velocity in the vicinity of $a_2$.
\par
The bottom panel shows the surface density at pericenter when we consider a size distribution from 1 to 100 km, with the initial number density of planetesimals proportional to $R^{-7/2}$.  Since $a_1$ varies significantly between different planetesimal sizes, we see multiple discrete spikes corresponding to individual planetesimal sizes.  For the system parameters considered here, planetesimals of all sizes between 1 and 100 km have almost the same value of $a_2$, so the peaks around at $a_2$ remain quite sharp.  The surface density of planetesimals in this simulation reaches a peak of 180 times its initial value, but again, we caution that diffusive processes not included in this simulation likely smooth the density profile somewhat.  The fact that $a_2$ is nearly the same for a wide range of planetesimal sizes is a consequence of the fact that planetesimals above a certain size orbit at nearly the forced eccentricity determined by the gravitational potential of the disk and binary \citep{Rafikov15b}.  It is therefore expected that in many cases the pile-up location will be nearly size-independent.

\subsection{Comparison with an axisymmetric disk}
\label{sect:Adachi}
While this paper has focused primarily on the implications of outwards radial drift, it is worth asking in general how much the drift rate differs from a simple extrapolation from the case of an axisymmetric disk with an eccentric planetesimal.  
\par
\begin{figure*}[htp]
\centering
\includegraphics[width = \textwidth]{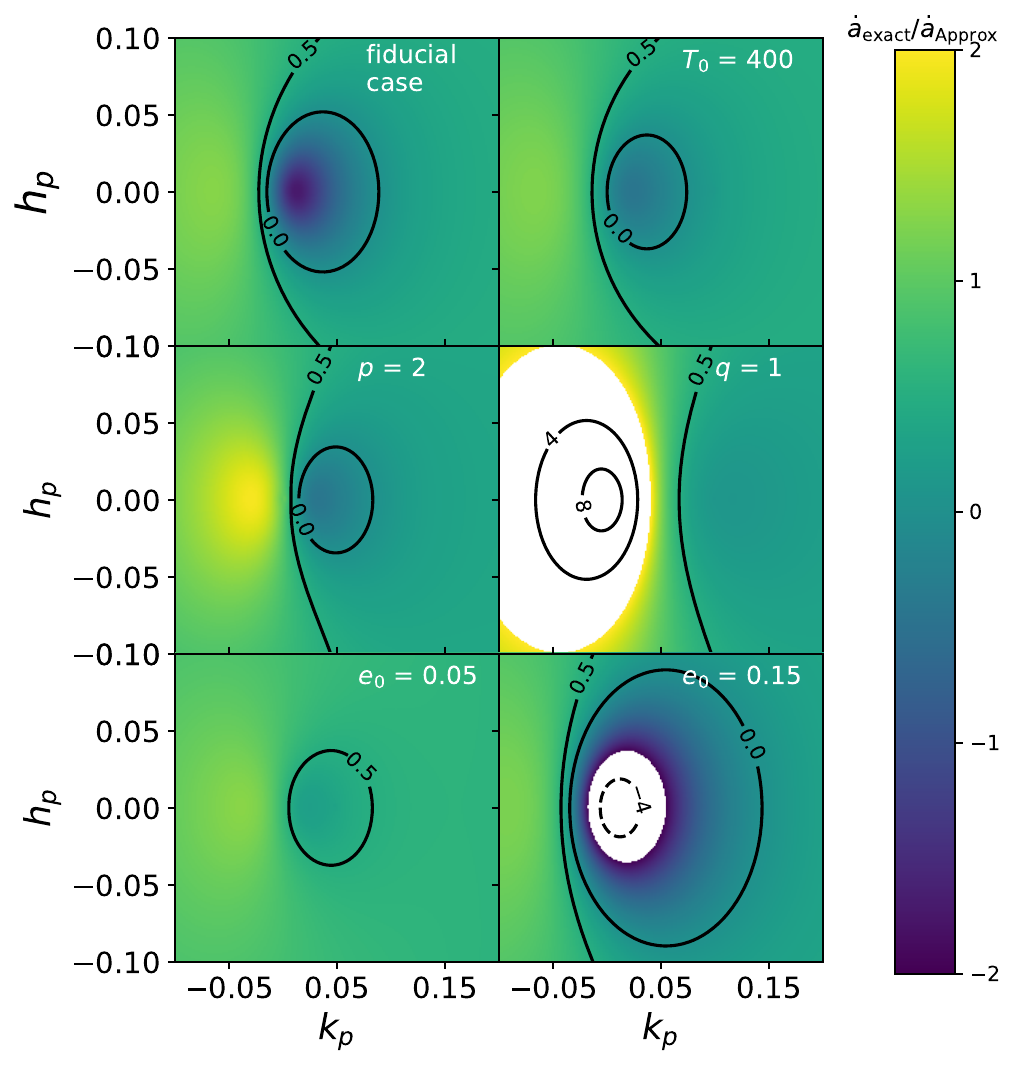}
\caption{Ratio of the inspiral rate given by Equations \eqref{eq:almostFinalInspiralEquation} and \eqref{eq:finalInspiralEquation} to that given by Equation 4.21 in \citet{Adachi76} (derived and intended by the authors to be used for an axisymmetric disk), with planetesimal eccentricity replaced by relative planetesimal-gas eccentricity as done in \citet{Rafikov15a}.  The ratio is shown by the color bar for values between -2 and 2, and with labeled contours for higher values.  We include also labeled contour lines at 0 and 0.5.}
 \label{AdachiInspiral}
\end{figure*}

We made this comparison for the disk models shown in Figure \ref{inspiralMap}.  Figure \ref{AdachiInspiral} shows the ratio of the inspiral rate calculated using Equations \eqref{eq:almostFinalInspiralEquation} and \eqref{eq:finalInspiralEquation}, to that calculated using Equation 4.21 of \citet{Adachi76} (intended to be used for an axisymmetric disk), with the eccentricity replaced with the relative eccentricity between the planetesimal and the gas streamline with the same semi-major axis, as done in \citet{Rafikov15a}.  In performing this approximate calculation, for a planetesimal of semi-major axis $a_p$, we approximated the gas density as equal to the gas density at pericenter for the streamline with semi-major axis $a_p$.  
\par
 Areas of major disagreement (i.e. areas in which the approximate treatment gets the sign wrong, or overestimates the rate by a factor of 10 or more) are present in all panels but one, but make up less than half of the area of the panel, except for the bottom right panel, in which such areas comprise almost exactly 50\% of the panel. 
\par
Looking at the bottom panels, we note that the amount of disagreement depends strongly on the disk eccentricity.  This makes sense, because the two methods agree as to the effect of the pressure gradient. For low eccentricity, the inspiral rate is dominated by the pressure gradient, and therefore the difference between the approximate and exact treatments is not large (see bottom left panel).  The most dramatic differences occur for the disk in which $e_0 = 0.15$ (bottom right panel), as in this case the eccentricity, rather than the pressure gradient, has the dominant effect on the particle-gas velocity.

\subsection{Effect of disk parameters}
\label{sect:diskParams}
 The radial drift rate is affected most strongly by the eccentricity profile of the disk.  For outward drift to be possible, depending on the temperature, the eccentricity must be increasing outwards and $\gtrsim 0.1$.  Simulations of disks in binary systems often find a region of the disk with these properties \citep[e.g][]{Marzari09, Martin20, Jordan21}, however the details vary substantially from one work to another.  \citet{Jordan21} ran a series of 2D radiative hydrodynamic simulations using using the FARGO code.  They found that the eccentricity profile was dramatically affected by choices regarding the simulation domain, the boundary condition at the inner edge of the disk, the treatment of thermodynamics, and numerical resolution.   \citet{Marzari09} found that inclusion of self-gravity substantially decreased the disk eccentricity, however even in their self-gravitating model with a perturber orbiting at 30 AU, the disk reached a maximum eccentricity of 0.25 at 1 AU; higher than in any of the models considered in this paper.  
\par
It is also of note that disks in many simulations precess.  For our high disk-mass case, precession, provided it is not too rapid, should not affect our results.  However, the case of a low mass-disk in which binary gravity is important (illustrated in Figure \ref{withPerturber}), outward radial drift is only possible in an aligned disk.  Rough apsidal alignment (though with substantial libration and some systematic offset) has been found in the simulations of \citet{Muller12, Gyergyovits14, Martin20}, however other simulations \citep{Paardekooper08, Marzari09, Marzari12} find anti-alignment or precession.
\par
In general, we find that the radial drift of planetesimals in  disks with eccentricity $\geq 0.1$ may show strong deviations from the values calculated using the approximate method in \citet{Rafikov15a}.  If these moderate eccentricities are combined with an outward-increasing eccentricity gradients in part of the disk, then there may be a local region in which the radial drift is outwards.  Although the true surface density and eccentricity profiles are unlikely to be well-represented by power laws, it seems plausible that for more realistic disk models, these conditions would be satisfied somewhere within the disk.
\par
The disk temperature is also important, since increasing the disk temperature by a given factor increases the pressure gradient and therefore $\eta$ by the same factor [see Equation \eqref{eq:eta}]. We use a disk temperature of 200 Kelvin at 1 AU, as appropriate for a disk heated by the radiation of the central star \citep{Chiang97}.  Models show that a massive disk undergoing viscous accretion will become several times hotter than this \citep{Dalessio98, Dullemond07}.  However, the relation between accretion rate and mid-plane temperature is not simple.  For example, some models suggest that accretion occurs mainly near the disk surface, with a low-viscosity ``dead" zone at the disk midplane.  In this case, the midplane temperature may be substantially cooler at a given accretion rate than in a simple model with vertically uniform viscosity \citep{Hirose11}.  More dramatically, \citet{Mori19} found that a magnetic wind driven accretion model can achieve high accretion rates with very low rates of mid-plane heating.  For example, they considered a disk with surface density $17,\!000$ g cm$^{-2}$ at 1 AU, with accretion rate of $8.2 \times 10^{-7} M_\odot/{\rm yr}$.  This resulted in a midplane temperature slightly under 200K, compared with $\sim 900$K in the viscous heating model with the equivalent accretion rate.  For our fiducial disk, the temperature must be increased to 600 K at 1 AU for there to be no value of the planetesimal eccentricity vector leading to outwards migration.  If we increase $e_0$ to 0.15 (as in the bottom right panel of Figure \ref{inspiralMap}), then this critical temperature increases to 1470 K.

\subsection{Effect of azimuthal pressure gradients}
\label{sect:pressureEffects}
Here, we justify the use of an azimuthally independent value for $\eta$ in Equation \eqref{eq:GasVelocity} by showing that pressure gradients along the streamline have a small effect on the velocity.  Starting from Equation \eqref{eq:PRhoSigma}, we use Bernoulli's law to calculate the change in velocity along a streamline:
\begin{equation}
\frac{v^2}{2} + \int_{P_p}^{P} \frac{d\tilde p}{\rho(\tilde p)} - \frac{GM}{r} =  {\rm constant} \rightarrow v = \sqrt{v_{\rm peri}^2  + \frac{2\gamma}{\gamma -1} \left[\frac{P_{\rm peri}}{\rho_{\rm peri}} - \frac{P}{\rho}\right] + 2GM \left(\frac{1}{r} - \frac{1}{a_g(1-e_g)}\right)}.
\label{eq:Bernoulliv}
\end{equation}
To isolate the effect of the azimuthal pressure gradient, we Taylor-expand the velocity given by Equation \eqref{eq:Bernoulliv} as
\begin{equation}
v = v_* + \Delta v,
\end{equation}
where $v_* = \sqrt{v_{\rm peri}^2 + 2GM\left(1/r - 1/[a_g(1-e_g)]\right)}$ is the velocity in the absence of a pressure perturbation along the streamline, and 
\begin{equation}
\Delta v = \frac{\frac{\gamma}{\gamma-1} \left[\frac{P_{\rm peri}}{\rho_{\rm peri}} - \frac{P}{\rho}\right]}{v_*}
\end{equation}
is the contribution due to the pressure gradient.  Using Equations \eqref{eq:finalSigma} and \eqref{eq:PRhoSigma}, and dropping all terms of order $e^2$, we find that the change in velocity between pericenter and apocenter, attributable to the pressure gradient along the streamline is 
\begin{equation}
\Delta v = \frac{2 \gamma \left[3-2q\right]}{\gamma + 1}e_g \left(\frac{c_s}{v_K(a_g)}\right)^2 v_K(a_g).
\end{equation}
This is less by a factor of $4 \gamma (3-2q)e_g/[(\gamma + 1)(p + (s+3)/2]$ with respect to the typical ``headwind" velocity due to the radial pressure gradient in a circular disk [see Equation \eqref{eq:eta}].  At location $a_0$ in our fiducial disk, this factor is 0.43 --- not a terribly small number.  However this analysis would suggest that by calculating $\eta$ at $a_g$ as we have done in Equation \eqref{eq:GasVelocity}, we are underestimating the gas velocity at apocenter, and overestimating it at pericenter, and thus these errors should partially cancel each other.  At any rate, it was shown in Figure \ref{inspiralMap} that our results are robust to a factor 2 increase in disk temperature above the fiducial value (corresponding to a factor 2 change in $\eta$), suggesting that this additional uncertainty will not alter any of the conclusions of the paper.

\subsection{Different treatments of disk thermodynamics}
Throughout this work, we assume that the gas in the disk behaves adiabatically on an orbital timescale, and that the disk height at a given azimuthal angle is given by $c_s r/v_k(r)$.  Here we explore three other limiting cases, and show that, except for the rather unrealistic assumption of an azimuthally invariant disk height along a given streamline, the effects on our results are minimal.  
\par
First, we consider an isothermal model in which the gas temperature along a streamline does not vary with azimuthal angle.  In this case, $\chi = 1$ [see Equation \eqref{eq:azimuthalVariations}], and re-derivation of Equation \eqref{eq:PRhoSigma} with this assumption yields
\begin{equation}
    \frac{\rho_g}{\rho_{\rm peri}} = \left(\frac{C^2}{D^3}\right)^{1/2}.
    \label{eq:isothermalrho}
\end{equation}
\par
Second, we consider a case in which the disk height is constant along a streamline.  In this case, the mid-plane density is directly proportional to the surface density, so
\begin{equation}
        \frac{\rho_g}{\rho_{\rm peri}} = C.
        \label{eq:constantHrho}
\end{equation}
This scenario is not entirely implausible as the rough timescale for the disk to come into vertical equilibrium ($H/c_s$) is of the same order as the dynamical timescale $r/v_K$.  However, as the time between apocenter and pericenter is $\pi$ times the dynamical timescale, it is likely a better approximation, as we have assumed in the rest of this paper, that the vertical structure of the disk is in local hydrostatic equilibrium.

\par
Finally, we consider a case in which the disk material is assumed to rapidly equilibrate its temperature in less than an orbital time.  In this case, the temperature anywhere along a streamline is given by Equation \eqref{eq:eccentricityAndSurfaceDensity} with $a$ replaced by $r$, resulting in the relation
\begin{equation}
    \frac{\rho_g}{\rho_{\rm peri}} = \frac{C}{D^{(3-s)/2}}.
    \label{eq:instantEqrho}
\end{equation}
\label{sect:diskThermo}
 \begin{figure*}[htp]
\centering
\includegraphics[width = \textwidth]{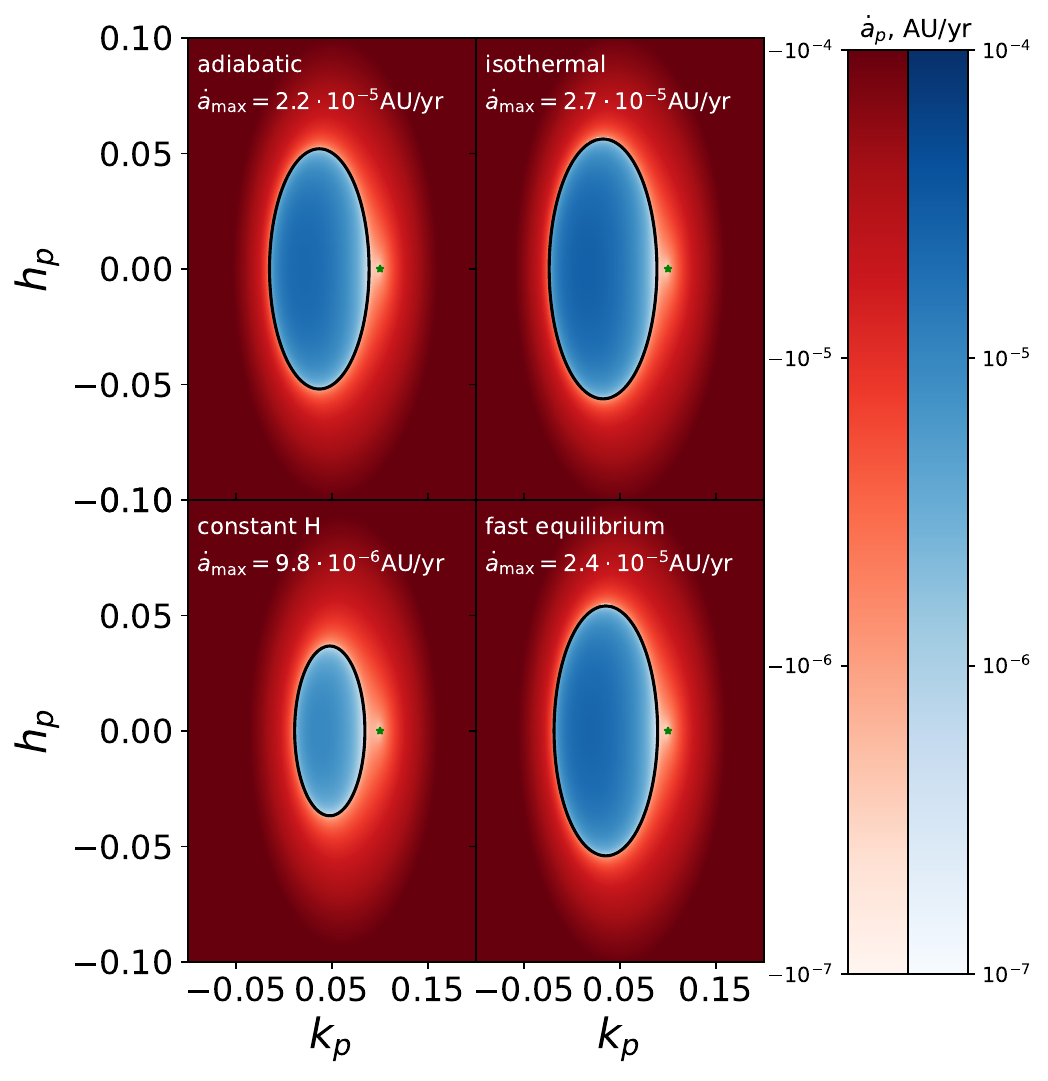}
\caption{The inspiral rate for a one kilometer planetesimal at 1 AU in a disk with our fiducial parameters, in the space of the components $k_p$ and $h_p$ of the planetesimal eccentricity vector.  Different panels correspond to different limiting cases of the disk thermodynamics as discussed in the text.  As in Figure \ref{inspiralMap}, the color shows the rate of radial drift, with red indicating drift towards the star, and blue indicating drift away from the star.  The peak (outwards) drift rate is labeled on each panel. The color bar saturates at $\pm 10^{-4}$ AU/yr. The green star in each panel represents the gas eccentricity.
}
 \label{thermoPlot}
\end{figure*}
Figure \ref{thermoPlot} shows the results of these different limiting cases for the disk thermodynamics.  The upper left panel shows our fiducial case of an adiabatic disk in local vertical hydrostatic equilibrium (and is therefore identical to the upper left panel of Figure \ref{inspiralMap}).   The upper right panel shows the case where we assume a constant temperature along a streamline [Equation \eqref{eq:isothermalrho}].  Compared with the adiabatic case, this results in a disk in which the variation of density along a streamline between pericenter and apocenter is higher, as the disk is warmer at apocenter.  Since in the disk models under consideration, an apsidally aligned planetesimal gains energy at pericenter and loses energy at apocenter, a disk with isothermal streamlines results in a slightly enlarged region of outward migration.
\par
In contrast, assuming a constant disk height along the streamline [Equation \eqref{eq:constantHrho}], significantly increases the ratio of the density at apocenter to that at pericenter compared with the fiducial case, and thus shrinks the region of outward migration, (see bottom-left panel of Figure \ref{thermoPlot}).  Finally assuming that material instantaneously equilibrates its temperature, and that the temperature is a sole function of $r$ [Equation \eqref{eq:instantEqrho}], results in a somewhat cooler temperature at apocenter, similar to the adiabatic case.  Except for the somewhat unrealistic case where we assume that disk height doesn't change along a given streamline the changes associated with assumptions about thermodynamics are minor.
\section{Conclusions}
\label{sect:conclusion}
In this paper, we calculated the rate at which planetesimals drift radially in eccentric disks under the influence of aerodynamic drag.  We found in some cases that the drift rates differ by an order of magnitude from those calculated for planetesimals in circular disks with the same eccentricity relative to the gas.  This is true even for relatively small ($\sim \! \!0.1$) disk eccentricities.  In contrast both with typical expectations in protoplanetary disks, as well as results of previous approximate calculations, we found that under some plausible assumptions about the disk eccentricity profile, it is possible that planetesimals may drift {\it away} from the central star.
\par
In addition to calculating radial drift as a function of planetesimal eccentricity, we used secular perturbation theory to calculate the equilibrium eccentricity as a function of planetesimal size for planetesimals evolving under the combined influence of gas drag, and gravitational perturbations from the companion star and the disk.  We verified that, as in the case of an axisymmetric disk, the eccentricity vector evolves much faster under the influence of gas drag than does the semi-major axis.  As a result, as planetesimals drift in the disk, they remain very close to their equilibrium eccentricity for their present semi-major axis.  We found two classes of model in which planetesimals orbiting with the equilibrium eccentricity experience outwards drift over a substantial range of orbital radii, suggesting that in circumstellar disks with a close companion star there will plausibly be regions in which the planetesimal migration is outwards. 
\par
The first such case is a low surface density disk apsidally aligned with the binary orbit.  In such a scenario, we found outward migration times of hundreds of thousands of years for a 1 km planetesimal, at a location where the disk eccentricity is 0.1.  Second, we studied planetesimals in a high-mass disk ($\sim \! \! 6$ times MMSN at 1 AU) at a location with the same value of the disk eccentricity.  Here, for favorable surface density and eccentricity profiles, we found outward migration timescales under 100,000 years, even for 30 km planetesimals.  In this case, the disk need not be aligned with the binary, and the disk mass itself is not critical, provided that it is large enough to dominate the gravitational perturbations of the binary at the location of the planetesimal.  
\par
The outer edge of the zone of outward migration is a stable attractor for planetesimals, and thus over time the surface density of planetesimals is enhanced near this location.  Sufficiently large planetesimals have their eccentricity vectors driven to the so-called ``forced eccentricity" determined by the gravitational perturbations.  Since the eccentricity vector is independent of planetesimal size, then the region of the disk in which planetesimals migrate outwards is also independent of planetesimal size.  Thus all planetesimals pile up at the same location.  In our model, this resulted in an enhancement by more than a factor of 100 in the planetesimal surface density at the edge of the zone of outward migration, though diffusive processes not accounted for in our modeling likely lower this number.

\bibliographystyle{apj}
\bibliography{eccentricDrift}

\counterwithin*{equation}{section}
\renewcommand\theequation{\thesection\arabic{equation}}
\appendix
\section{Simplified model}
\label{append:simpleModel}

Equations \eqref{eq:almostFinalInspiralEquation} and \eqref{eq:finalInspiralEquation} provide a recipe to numerically calculate the inspiral rate, but do not offer a lot of insight into its behavior.  In this appendix (which can be safely skipped by a reader interested only in the results and implications of this work), we aim to provide more insight into the behavior of the radial drift rate.  We derive an analytic estimate of the inspiral rate of an apsidally aligned planetesimal by taking an average of the instantaneous $\dot a_p$ at pericenter and apocenter weighted by the respective dynamical timescales.  This does not provide a very accurate answer, but gives an understandable explanation for the trends shown in Figure \ref{inspiralMap}.   We assume for simplicity that the planetesimal semi-major axis $a_p = a_0$.  Since the relevant quantities are power-laws in semi-major axis, this does not result in any loss of generality, because for any value of $a_p$, we can set $a_0 = a_p$, and rescale $\Sigma_0$, $e_0$ and $T_0$ as appropriate.  
\par
We characterize the planetesimal eccentricity in terms of a small parameter $\epsilon$, defined as
\begin{equation}
\epsilon = e_0 - e_p.
\end{equation}
 We calculate the planetesimal and gas velocities at the planetesimal's pericenter and apocenter, considering terms proportional to $\eta_0^2$, $\eta_0\epsilon$, and $\epsilon^2$, but no higher-order terms.  $e_0$ is not treated as a small parameter.  To begin, we use Equation \eqref{eq:EccentricVelocities} to calculate the planetesimal velocity at pericenter and apocenter, to first order in $\epsilon$:
\begin{equation}
v_p^{\rm peri} = v_0^{\rm peri} \left(1 - \frac{\epsilon}{1-e_0^2}\right); \quad \quad v_p^{\rm apo} = v_0^{\rm apo} \left(1 + \frac{\epsilon}{1 - e_0^2} \right),
\label{eq:particleExtremalVelocities}
\end{equation}
where
\begin{equation}
v_0^{\rm peri} = \sqrt{\frac{GM (1 + e_0)}{a_0 (1 - e_0)}}; \quad \quad v_0^{\rm apo} = \sqrt{\frac{GM (1 - e_0)}{a_0 (1 + e_0)}}.
\end{equation}
  We can then write, to first order in $\epsilon$, the semi-major axis $a_g^{\rm peri}$ of the gas streamline at the location of the planetesimal's pericenter, and $a_g^{\rm apo}$ of the gas streamline at the location of the planetesimal's apocenter:
\begin{equation}
a_g^{\rm peri} = a_0 \times (1 + \alpha_{\rm peri} \epsilon); \quad \quad a_g^{\rm apo} = a_0 \times (1 + \alpha_{\rm apo} \epsilon),
\label{eq:diska}
\end{equation}
where
\begin{equation}
\alpha_{\rm peri} = \frac{1}{1 + e_0(q-1) }; \quad \quad \alpha_{\rm apo} = \frac{-1}{1 + e_0 (1 - q)}.
\label{eq:alpha}
\end{equation}
We note that Equations \eqref{eq:diska} are not valid in the case that $\epsilon > e_0$, since in this case, the locations of pericenter and apocenter switch places, i.e. the planetesimal goes from being aligned with the disk to being anti-aligned with the disk.
\par
In combination with Equation \eqref{eq:eccentricityAndSurfaceDensity}, this enables us to write the eccentricity of the gas streamline intersecting the planetesimal pericenter and apocenter, again to first order in $\epsilon$:
\begin{equation}
e_g^{\rm peri} = e_0 (1 - q \alpha_{\rm peri} \epsilon); \quad \quad e_g^{\rm apo} = e_0  (1 - q \alpha_{\rm apo} \epsilon).
\label{eq:ed}
\end{equation}
Then, using Equations \eqref{eq:GasVelocity}, \eqref{eq:diska}, \eqref{eq:alpha} and \eqref{eq:ed}, we can express the {\it gas} velocity at the particle's pericenter and apocenter to first order in $\epsilon$ and $\eta_0$:
\begin{equation}
v_g^{\rm peri} = v_0^{\rm peri} \left[1 - \eta_0 - (\alpha_{\rm peri} \epsilon) \times \left(\frac{1}{2} + \frac{q e_0}{1 - e_0^2} \right) \right]; \quad \quad v_g^{\rm apo} = v_0^{\rm apo} \left[1 - \eta_0 - \alpha_{\rm apo} \epsilon \left(\frac{1}{2} - \frac{e_0q}{1 - e_0^2}\right)\right].
\label{eq:GasExtremalVelocities}
\end{equation}
Combining Equations \eqref{eq:particleExtremalVelocities} and \eqref{eq:GasExtremalVelocities}, we can write the relative particle-gas velocity at pericenter and apocenter as 
\begin{equation}
v_{\rm rel}^{\rm peri, apo} = v_K(a_0) \left[c_{\rm peri, apo} \eta_0 + d_{\rm peri, apo} \epsilon \right],
\label{eq:vrelShort}
\end{equation}
where 
\begin{equation}
c_{\rm peri} = \sqrt{\frac{1 + e_0}{1 - e_0}}; \quad \quad d_{\rm peri} = \frac{e_0-1}{2 \sqrt{1-e_0^2} \left[1 + e_0(q-1)\right]},
\label{eq:cp}
\end{equation}
\begin{equation}
c_{\rm apo} = \sqrt{\frac{1 - e_0}{1 + e_0}}; \quad \quad d_{\rm apo} = \frac{1 + e_0}{2 \sqrt{1-e_0^2} \left[1-e_0(q-1)\right]}.
\label{eq:ca}
\end{equation}
Since we are only considering terms proportional to $\epsilon^2$, $\epsilon \eta_0$, and $\eta_0^2$, we ignore the dependence of the gas density at pericenter and apocenter on $\epsilon$, as its inclusion would not affect the value of any of these terms in the expression for $\dot a_p$. This gives us, from Equations \eqref{eq:pericenterDensity} and \eqref{eq:PRhoSigma}, 
\begin{equation}
\rho_g^{\rm peri} = \rho_0; \quad \quad \rho_g^{\rm apo} = \rho_0 \left(C_{\rm apo}^2/D_{\rm apo}^3 \right)^{1/(\gamma + 1)},
\label{eq:extremalDensity}
\end{equation}
where 
\begin{equation}
\rho_0 = \frac{\Sigma_0}{\sqrt{2\pi} H(a_0(1-e_0))},
\end{equation}
[see Equation \eqref{eq:pericenterDensity}] and
\begin{equation}
C_{\rm apo} = \frac{1 - e_0^2 + qe_0(1 + e_0)}{1 - e_0^2 + q e_0(-1 + e_0)}; \quad \quad D_{\rm apo} = \frac{1 + e_0}{1 - e_0}.
\end{equation}
For the fiducial values $e_0 = .1$, $q = -1$ and $\gamma = 10/7$, we find that $\rho_g^{\rm apo} = 0.66 \rho_g^{\rm peri}$.
\par
Inserting Equations \eqref{eq:vrelShort} and \eqref{eq:extremalDensity} into \eqref{eq:dragForce}, yields the following for the drag force at pericenter and apocenter:
\begin{equation}
F_d^{\rm peri} = - \frac{\pi s_{\rm peri} C_dR^2 \rho_0 v_K^2(a_0)}{2} \left(c_{\rm peri} \eta_0 + d_{\rm peri} \epsilon \right)^2;
\label{eq:pericentralForce}
\end{equation}
\begin{equation}
F_d^{\rm apo} = - \frac{\pi s_{\rm apo} C_d R^2 \rho_0 v_K^2(a_0) \left(C_{\rm apo}^2/D_{\rm apo}^3 \right)^{1/(\gamma + 1)} }{2} \left(c_{\rm apo} \eta_0 + d_{\rm apo} \epsilon \right)^2,
\label{eq:apocentralForce}
\end{equation}
where $s_{\rm peri, apo} = {\rm sgn}(v_{\rm rel}^{{\rm peri, apo}})$.
\par
Then, we define the dynamical timescales of the planetesimal at pericenter and apocenter, again ignoring the $\epsilon$ dependence, as that does not affect our solution at the desired order:
\begin{equation}
\tau_{\rm peri} = \frac{r_{\rm peri}}{v_{\rm peri}} = \sqrt{\frac{a_0^3 (1-e_0)^3}{GM(1+e_0)}}; \quad \quad \tau_{\rm apo} = \frac{r_{\rm apo}}{v_{\rm apo}} = \sqrt{\frac{a_0^3 (1+e_0)^3}{GM(1-e_0)}}.
\end{equation}
We then approximate the inspiral rate as 
\begin{equation}
\langle \dot a_p \rangle \approx \frac{\dot a_{\rm peri} \tau_{\rm peri} + \dot a_{\rm apo} \tau_{\rm apo}}{\tau_{\rm peri} + \tau_{\rm apo}},
\label{eq:inspApprox}
\end{equation}
 where $\dot a_{\rm peri}$ and $\dot a_{\rm apo}$ are the values of $\dot a_p$ evaluated at pericenter and apocenter respectively.  Using Equations \eqref{eq:almostFinalInspiralEquation}, \eqref{eq:pericentralForce} and \eqref{eq:apocentralForce} to calculate the $\dot a_{\rm peri}$ and $\dot a_{\rm apo}$, and inserting the results into Equation \eqref{eq:inspApprox} gives
\begin{align}
& \langle \dot a_p \rangle = \frac{-\pi C_dR^2\rho_0a_0v_K(a_0) (1-e_0)}{m_p \left(\sqrt{\frac{(1-e_0)^3}{1+e_0}} + \sqrt{\frac{(1+e_0)^3}{1-e_0}}\right)} 
\nonumber \\
& \times \left[\eta_0^2 (s_{\rm peri} c_{\rm peri}^2 + \psi s_{\rm apo} c_{\rm apo}^2) + 2 \eta_0 \epsilon (s_{\rm peri} c_{\rm peri} d_{\rm peri} + \psi s_{\rm apo} c_{\rm apo} d_{\rm apo}) + \epsilon^2 (s_{\rm peri} d_{\rm peri}^2 + \psi s_{\rm apo} d_{\rm apo}^2) \right],
\label{eq:finalAnalytic}
\end{align}
with
\begin{equation}
\psi = \left( \frac{C_{\rm apo}^2}{D_{\rm apo}^3}\right)^{\frac{1}{\gamma + 1}} \times \left(\frac{1+e_0}{1-e_0}\right).
\end{equation}
With this analysis in mind, we can revisit the trends shown in Figure \ref{inspiralMap}.  
 \begin{figure*}[htp]
\centering
\includegraphics[width = \textwidth]{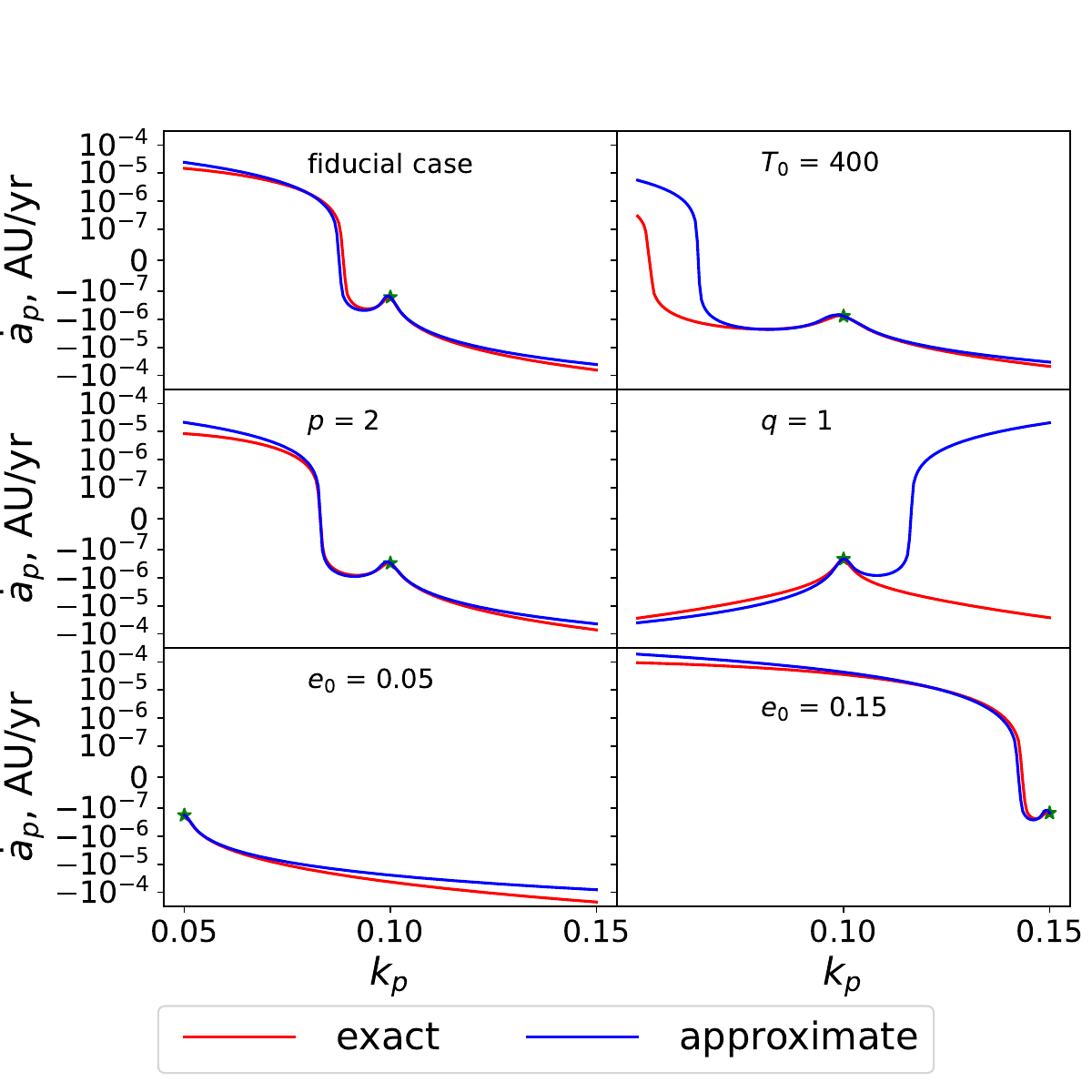}
\caption{This shows the inspiral rate, for planetesimals apsidally aligned with the gas streamlines ($h_p = 0, k_p \geq 0$).  The red curves are made using the exact solution given by the method described in Section \ref{sect:exactCalc}.  The blue curves are from the analytic approximation in Equation \eqref{eq:finalAnalytic}.  Each panel corresponds to the same set of disk parameters as in the corresponding panel of Figure \ref{inspiralMap}.  The green star corresponds to $k_p = e_0$, or $\epsilon = 0$.
}
 \label{analyticComparison}
\end{figure*}
Figure \ref{analyticComparison} shows the inspiral rate for apsidally aligned planetesimals ($h_p = 0$, $k_p \geq 0$) for the same disk parameters as in Figure \ref{inspiralMap}.  The red curves show the exact solution given by Equations \eqref{eq:almostFinalInspiralEquation} and \eqref{eq:finalInspiralEquation}, and the blue curves show the approximate solution given by Equation \eqref{eq:finalAnalytic}.  In the vicinity of the green star; i.e. for $\epsilon$ close to zero, the exact and approximate solutions match well.  In this case, the blue curves are parabolic, opening downwards. 
\par
Although this is difficult to see on the plots, the vertex is very slightly displaced from $\epsilon = 0$.  As can be seen from Equation \eqref{eq:finalAnalytic}, the vertex would be exactly at $\epsilon = 0$, if $\eta_0 = 0$.  In the fiducial case, the vertex is located at $\epsilon = 4.7\times 10^{-4} \rightarrow e = .09953$.  Since $d_{\rm peri}$ is negative in the fiducial case, then $s_{\rm peri}$ must change sign as we go to higher values of $\epsilon$, and terms proportional to $\epsilon^2$ dominate over those including $\eta_0$.  Furthermore, since $\psi < 1$, and $|d_{\rm peri}| < |d_{\rm apo}|$, it is clear that for large values of $\epsilon$, the radial drift becomes positive according to Equation \eqref{eq:finalAnalytic}.  However, this may not happen in practice if the $\eta_0$ term is large enough to dominate over the range of $\epsilon$ for which Equation \eqref{eq:finalAnalytic} is a valid approximation.  As previously noted, in addition to the neglect of higher-order terms, for $\epsilon > e_0$, Equations \eqref{eq:diska} are no longer valid, even to order $\epsilon^2$.  Therefore, outward directed radial drift is unlikely to occur in a disk with eccentricity less than several percent (barring a very sharp eccentricity or surface density gradient), as it will be overwhelmed by the fact that the radial pressure gradient drives inwards drift.  
\par
The results presented in this appendix, while useful for building intuition, are not sufficiently accurate to be used for quantitative calculations, and as can be seen from Figure \ref{analyticComparison}, deviations on the order of a factor of 2 occur even for values of $\epsilon$ of a few percent.  This could be because only considering the effect at pericenter and apocenter is not valid.  However, as the curves all converge around $\epsilon = 0$, we do not believe that this is the main source of error.  Instead, as there is a near cancellation at pericenter and apocenter, we believe that the coefficient of term proportional to $\epsilon^2$ in Equation \eqref{eq:finalAnalytic} is very small compared with the coefficient of the higher-order terms, and therefore the approximation breaks down at small $\epsilon$, even getting the sign wrong for part of the panel corresponding to $q = 1$.

\end{document}